\documentclass[secnumarabic,amssymb, nobibnotes, aps, prd]{revtex4}%
\usepackage{graphicx}
\usepackage{dcolumn}
\usepackage{bm}
\usepackage{amsmath}%
\usepackage{amsfonts}%
\usepackage{amssymb}
\begin{document}

\title{Amplified Doppler shift observed in diffraction images as function of the COBE ``ether drift'' direction}

\author{Carlos. E. Navia$^*$}
\address{Instituto de F\'{\i}%
sica Universidade Federal Fluminense, 24210-130,
Niter\'{o}i, RJ, Brazil} 

\author{Carlos. R. A. Augusto}
\address{Instituto de F\'{\i}%
sica Universidade Federal Fluminense, 24210-130,
Niter\'{o}i, RJ, Brazil}

\date{\today}

\begin{abstract}
We report results on an ``one-way light path'' laser diffraction experiment as a function of the laser beam alignment relative to the Earth's velocity vector obtained by COBE measurements of the Doppler shift in the cosmic microwave background radiation (CMBR). An amplified Doppler shift is observed in the diffraction images, and the effect is
compatible with a ``dipole'' speed of light anisotropy due to Earth's motion relative to the ``CMBR rest frame'', with an amplitude of $\delta c/\bar{c}=0.00123$. This amplitude coincides with the value of the dipole temperature anisotropy $\delta T/\bar{T}=0.00123$ of the CMBR obtained by COBE. Our results point out that it is not possible to neglect the preferred frame imposed by the cosmology and they are well described by the Ether Gauge Theory (an extension of the Lorentz's ether theory) and it satisfies the cosmological time boundary condition.

\end{abstract}


\maketitle

\section{Introduction}

The Special Relativity Theory (SRT) introduces an universal fundamental constant, the speed of light in space, it is constant in all frames. Hence all reference frames are completely equivalent and an Ether rest frame is superfluous. 
The experimental basis of the SRT has been pointed out by T. Roberts \cite{roberts00}, where a lot of experimental tests done by many authors are summarized. Following this article we found several experiments about round-trip test of light speed isotropy (two-way light path), from the traditional Michelson-Morley to modern Laser/Maser tests, as well as, about one-way light path test of light speed isotropy. In all cases, the experiments have been mounted in Earth's surface on a rotating table or fixed to the Earth, in this last case it is looked for effects due to the Earth's rotation. Only the upper limits have been found and they are in agreement with a null effect, all results confirm the prediction of the SRT, or in other words, the SRT has not been refuted by any experiment.  

However, recently there are evidences suggesting that the propagation of light over cosmological distances has anisotropy characteristics\cite{nodland97}, this means that the speed of light is not a true constant, it
depends on direction and polarization. This is a further indication in favor of the existence of a preferred reference frame. This picture is in agreement with the interpretations of the COBE\cite{smoot91} measurements giving the Earth's ``absolute'' velocity in relation to the uniform cosmic microwave background radiation(CMBR). Of course, there are also interpretations claiming that the COBE measurements only give a velocity for the ``relative'' motion between the Earth and the CMBR \cite{yaes93}. For instance, it is possible to remove the Earth motion to obtain a ``virtual'' image, where an isotropic distribution of CMBR with small fluctuations ($\delta T/T \sim 10^{-5}$) can be seen.

On the other hand, the Big Bang cosmology is well described by the Robertson-Walker-Friedmann metric\cite{hans90}, the time in this metric called as the ``cosmological time'' has an absolute beginning and it coincides with the begin of the Big Bang, in other word, it is possible to estimate the absolute age of the universe. In short,
the COBE measurements of the Earth motion relative to the ``CMBR rest frame'' on an inflationary scenario, described by the Robertson-Walker-Friedmann metric, lead to the concept of absolute space-time, because in this scenario it is not possible to remove the Earth's motion without altering the Robertson-Walker-Friedmann metric\cite{walker}.

The consequences of an absolute space-time assumption are the breakdown of the symmetry-energy and symmetry-momentum
(linear and angular) conservation laws. The extension of the effects of the violation of these symmetries is still not known. So far, we know that probably they are relevant only in cosmological scale. In fact, in 1999, Coleman and Glashow began to wonder, if Lorentz's symmetry is certainly an exact symmetry of the nature, in the region close to the
Plack's scale energy. The Lorentz symmetry violation may be correlated to the existence of an absolute reference frame created by the Big-Bang.

The main objective of this paper is to present experimental results on an ``one-way light path'' laser diffraction  experiment mounted in the shell of the TUPI muon telescope \cite{navia05} and that shows clearly that the speed of light depends on the propagation direction. The effect is observed as a amplified Doppler shift in the diffraction images as function of the laser beam alignment relative to the Earth's velocity vector.

 The experiment is an improved version about light diffraction suggested by C. M. G. Lattes \cite{lattes80} in 1980. The results are well described by the the so called Ether Gauge Theory (EGT), and it is an extension of the Lorentz's ether theory. The gauge of the length is:  the moving rods undergo the shortening by Lorentz factor and the gauge of the time is: the moving clocks undergo the slowdown by Lorentz factor. The EGT scenario
satisfies the cosmological time boundary condition (see next section) and leads naturally to a non-isotropic ``one-way'' light propagation relative to the Earth frame and where there is a preferred direction.

The assumption of a preferred frame agree also with a previous analysis made by Brans and Stewart \cite{brans73} where a description of the topology of the universe has imposed a preferred state of rest so that the principle of special relativity, although locally valid, is not globally applicable. The analysis of The Global Positioning System (GPS) carried out by Hatch \cite{hatch02} provides also strong indirect evidence for the presence of an ether-drift velocity.

\section{Ether Gauge Theory for light propagation}

We presents here a straightforward analysis about light propagation on the basis of the Ether Gauge Theory (EGT)
formalism based upon a modification of the Lorentz ether theory in order to include the cosmological time constraint. This exigency is satisfied by Tangherlini \cite{tangherlini61} and Selleri \cite{selleri96} transformation.
This scenario admit an absolute referential frame $\Sigma_H$ called hereafter as the Hubble's frame \cite{lattes80} and it is linked with the ``CMBR rest frame''. The Hubble's frame is defined as:\\
(a) The frame where the CMBR is isotropic and \\
(b) the frame where light propagation is isotropic.

For simplicity and without loss of generality, only a bi-dimensional analysis is made. 
The Tangherlini-Selleri transformation from $\Sigma_H$ to $\Sigma_E$ in motion relative to $\Sigma_H$ with velocity parameter $\beta$ and where we choose the $x_H$ and $x_E$ as the direction of the relative motion has been found to be
\begin{equation}
x_E=\gamma(x_H-Vt_H), \;\; y_E=y_H \;\;and\;\;t_E=\gamma^{-1}t_H,
\end{equation}
where $\gamma=(1-\beta^2)$. 
The main difference between the EGT transformation (Tangherlini-Selleri transformation) and the SRT transformation (Lorentz transformation), constitutes the transformation of the time, the Tangherlini-Selleri time transformation satisfied the cosmological constraint (synchronization of clocks), if $t_H=0$ then $t_E=0$. The main consequence of this scenario is that, in the $\Sigma_E$ frame, the axis-x  (direction of $V$) behaves as a preferred direction.

The transformation equations for the velocity of a moving point can be found differentiating Eq.(1) with respect to $t_E$. 
\begin{equation}
\frac{dx_E}{dt_E}=V_{Ex}=\gamma^2(V_H \cos \theta_H-V), \;\;\\
\frac{dy_E}{dt_E}=V_{EY}=\gamma(V_H\sin \theta_H ).
\end{equation}
Now, if we denote
$c_{Ex}=V_{Ex}$, $c_{Ey}=V_{Ey}$ and $c_H=V_H$, the speed of light components in the $\Sigma_E$ frame are
\begin{eqnarray}
c_{EX}=c_H\;\gamma^2(\cos \theta_H-\beta),\;\;\;
c_{EY}=c_H\;\gamma \sin \theta_H,
\end{eqnarray} 
and the module of this velocity can be obtained as
\begin{equation}
c_E=\sqrt{c_{EX}^2+c_{EY}^2}=c_H\;\gamma^2(1-\beta \cos \theta_H).
\end{equation}
The last equation means that in the $\Sigma_E$ frame the speed of light is not a constant, but it depends on $\theta_H$.
Fig.1 shows the space-time diagram (light cone) in the $\Sigma_E$ frame. It is possible to see
that the axis-X  (direction of $V=V_E$) behaves as a preferred direction. The cone of light is symmetric in relation to the Y-axis, while it is asymmetric in relation to the X-axis. The slope of the lines is proportional to the inverse value of the speed of light. For illustrative purpose,
a large value for $\beta(=0.3)$ in this diagram has been used. The effect, for instance, due to the Earth motion relative to the CMBR rest frame is tiny ($\beta=0.001237$). From this picture, it is possible to see that there is a resulting vector speed of light, and it is represented as $V_e$ in Fig.1, and hereafter it will be called as the emission vector. The emission vector points in opposite direction of the vector $V_E$ and, following Eq.(4), the emission vector has a module given by
\begin{equation}
\left|V_e\right|=c_H\gamma^2(1+\beta)-c_H\gamma^2(1-\beta)=2c_H\gamma^2\beta,
\end{equation}
with amounts to $\left|V_e\right|\approx 2 \times \left|V_E\right|$. The effect due to the Earth motion gives
$\left|V_e\right|\approx 742\;km\;s^{-1}$.

On the other hand, the one-way (forward and backward) speed
of light in the Earth frame ($\Sigma_E$) are defines as
\begin{eqnarray}
c_E^f=c_H\;\gamma^2[1-\beta \cos (\theta_H)],\\
c_E^b=c_H\;\gamma^2[1-\beta \cos (\theta_H+\pi)],
\end{eqnarray}
Under this conditions, the one-way speed of light ``dipole'' anisotropy is defined as
\begin{equation}
\frac{\delta c_E}{\overline{c_E}}=\frac{c_E^f - \overline{c_E}}{\overline{c_E}},
\end{equation}
where $\overline{c_E}$ is the average values as 
\begin{equation}
\overline{c_E}=\frac{c_E^b + c_E^f}{2},
\end{equation}
with Eq.(6) and Eq.(7), Eq.(8) becomes   
\begin{equation}
\frac{\delta c_E}{\overline{c_E}}=-\beta \cos\theta_H,
\end{equation}
Thus the EGT scenario predicts a ``dipole'' speed of light anisotropy with an amplitude $\delta c_E/\overline{c_E}=\beta(=0.001237)$, which coincides with the value of the dipole temperature anisotropy ($\delta T/\bar{T}=0.001237$) of the CMBR as observed by COBE. 

Fig.2 summarized the situation, where the dependence of the ``dipole'' speed of light anisotropy is plotted as function of $\theta_H$. Following Fig.2, it is possible to see that in Earth-based experiments, the choice of the direction of the light beam is essential, An random choice for the light beam direction, as in experiment mounted on a rotating table, can easily lead to ambiguous results. In order to compensate the change of direction due to the Earth rotation, the Earth-based experiments  must be mounted like a optical telescope using an equatorial ensemble, for intance.

\section{The TUPI laser diffraction experiment}

The TUPI muon telescope is installed on the campus of the Universidade Federal
Fluminense, Niter\'oi, Rio de Janeiro-Brazil at sea level. The position is: latitude:
$22^{0}54^{\prime}33^{\prime\prime}$ S, longitude: $43^{0}08^{\prime}39^{\prime\prime}$ W. The TUPI muon telescope has an equatorial
assembly and a servo-mechanism which allows the axis of the telescope 
to be pointed so as to accompany a given source \cite{navia05,augusto05,naviab05}. 

On the other hand, the bigger temperature anisotropy in the cosmic background radiation is represented by a dipole with an amplitude of $\delta T/T=1.23\times 10^{-3}=0.123\%$ which arises from
the motion of the solar system barycenter with a velocity $V_E=371\pm 0.5kms^{-1}$ ($\beta=0.001237\pm 0.000002$) at 68\%CL, relative to the so called ``CMBR rest frame'' and towards a point whose equatorial coordinates are
$(\alpha,\;\delta)=(11.20^h\pm0.01^h,\;-7.22^0\pm 0.08^0$)\cite{smoot00}. This direction points for the Crater constellation. 

A laser light diffraction experiment has been mounted in the shell of the TUPI muon telescope, as is represented in Fig.3. The layout of the diffraction device is shows in Fig.4, the experiment permits us to obtain the variations of the diffraction line positions, $Y_n$, as a function of the laser beam alignment, $\theta_H$, relative to the Earth's velocity vector, obtained by COBE. 

This experiment is an improved version of an experiment accomplished in 1980 in Campinas Brazil and suggested by 
C. M. G. Lattes (with unpublished results). Just as in the modern version of the Michelson-Morley experiment \cite{muller03}, where anisotropies are sought in the light (radiation) propagation due the Earth rotation, the Lattes version tried to measure variations of the positions ($Y_n$)(see Fig.4) of the diffraction lines due to the Earth's rotation. This requires to make measurements over long time of at least several days to weeks, and it is necessary to take into account the atmospheric pressure and temperature variations among others, because the daily atmospheric variations may mask the effect. 

In the TUPI laser diffraction experiment, a complete set of measurements is made in only ten minutes.
In addition, just as it is shown below, the $Y_n$ variations are bigger when the direction of the laser beam goes from parallel to anti-parallel to the Earth's velocity vector (ether drift). 

Fig.4 shows a schematic diagram of the TUPI laser diffraction experiment in the Earth ($\Sigma_E$ frame).
Following this figure, it is possible to see that the maxima of intensity of the diffraction images satisfy the condition
\begin{equation}
\sin \alpha_n=\pm n\frac{\lambda_E}{\Delta},\;\;\;with\;\;(n=0,1,2,....),
\end{equation}
where $\lambda_E$ is the wave length in the $\Sigma_E$ frame and $\Delta$ is the diffraction grating step. Under the conditions $\sin\alpha_n \cong \tan\alpha_n$ we have
\begin{equation}
Y_n=d_0\;\tan \alpha_n \approx\pm n\frac{d_0}{\Delta}\lambda_E.
\end{equation}
The wave length $\lambda_E$ can be obtained as the ratio between the speed of light $c_E$ and the light frequency
$\nu_E$ resulting into $\lambda_E=c_E/\nu_E$. An expression for the $\lambda_E$ as a function
of the angle $\theta_H$ can be obtained using the transformation equations from $\Sigma_E$ to $\Sigma_H$ supplied by the EGT formalism, which for $c_E$ gives
\begin{equation}
c_E=c_H\;\gamma^2(1-\beta \cos\theta_H).
\end{equation}
On the other hand, according to the EGT formalism (see section 2), the space-time diagram or light cone in the $\Sigma_E$ frame is asymmetric in relation to the axis parallel to the Earth's speed vector, $V_E$, and there is the emission's vector (see Fig.1) represented as $V_e$ in Fig.4, where P represent the point of emission of light in the laser. This vector points in opposite direction of the Earth's motion and its module is $\left|V_e\right|\cong 2\times\left|V_E\right|$. This is equivalent to consider a source moving away at speed $V_e=2\times V_E$.
While in the $\Sigma_H$ frame we have a source moving away at speed $V_e\cong 2\;V_E-V_E=V_E$.

Consequently, a relation for the Doppler effect must be used to obtain $\nu_E$. The relation for the
Doppler effect obtained on the basis of the EGT formalism is the same as the relation obtained by using the SRT one since they are in agreement with experimental measurement \cite{ives38}. This suggest that locally, both 
the SRT and the EGT theories are indistinguishable, although they differ when global aspects are considered. The relation of the Doppler effect is 
\begin{equation}
\nu_E=\frac{\sqrt{1-\beta^2}}{(1-\beta\cos\theta_H)}\nu_H.
\end{equation}

Including the last two expressions in $\lambda_E$ result in
\begin{equation}
\lambda_E=\frac{(1-\beta\cos\theta_H)^2}{(1-\beta^2)^{3/2}}\lambda_0,
\end{equation}
where $\lambda_0=c_H/\nu_H$. Thus the combined effect, the $c_E=c_E(\theta_H)$ and the apparent source motion
$\nu_E=\nu_E(\theta_H)$ results in a ``amplified'' Doppler shift of the diffraction images. Expanding in terms of $\beta$ we have
\begin{equation}
\lambda_E=\left[1-2\cos\theta_H\;\beta+(\frac{3}{2}+\cos\theta_H^2)\;\beta^2-3\cos\theta_H\;\beta^3+O(\beta^4)\right]\lambda_0.
\end{equation}
Considering only the first order terms in $\beta$, $Y_n$ reads
\begin{equation}
Y_n\cong\pm \frac{n\;d_0}{\Delta}\lambda_0\left[1-2\cos\theta_H\;\beta\right].
\end{equation}
Now, if we denominate $\overline{Y_n}=n\;d_0 \lambda_0/\Delta$, the relative variation of $Y_n$ can be found as
\begin{equation}
\frac{Y_n-\overline{Y_n}}{\overline{Y_n}}=\frac{\Delta Y_n}{\overline{Y_n}}=-2\cos\theta_H\;\beta.
\end{equation} 
 Thus the EGT scenario for the relative variation (shift) $\Delta Y_n/Y_n$ of a diffraction image predicts an amplitude twice bigger than the dipole speed of light anisotropy and differs from the SRT prediction where $\Delta Y_n/Y_n=0$.    
We have compared these two predictions with experimental data obtained from
measurements on light diffraction produced by a laser beam in perpendicular incidence on a diffraction grating as a function of the laser beam alignment relative to the Earth's velocity vector. 

So far, we have three complete sets of independent observations made in three days, February 20 and 21 and March 9 of 2006. In order to avoid contaminations on the diffraction observations due to atmospheric fluctuations such as pressure and temperature among others, each set of observations were obtained in ten minutes by 
photographing the diffraction image in 19 different positions of the angle $\theta_H$, from $\theta_H=0^0$ up to $\theta_H=180^0$ in steps of $10^0$. A digital machine of 3.2 Mega pixels, fixed in the telescope has been used. An off-line analysis of the digital images has been done, by using a special software. The method permits us to count the number of pixels between two points of the image. Our goal is to obtain only the relative variations in the position of the diffraction images, and it is not necessary to do calibrations. 

We present here the measurements made on March 9, 2006 where a diffraction grating of $600\;lines/mm$
and a Helio-Neon Laser of $632.8\;nm$ and $1.2mW$ has been used. Each position $Y_n$ has been measured with an accuracy of $\sim 99.87\%$. Fig.5 shows the amplified Doppler shift observed through the overlap of diffraction images as a function of the laser beam alignment relative to the Earth's velocity vector The images were filtered and amplified using always the same criteria. The colors in the images are artificial and the vertical lines were placed only as a visual guide. A quantitative result, showing the relative amplified Doppler shift as function of the laser beam alignment, is summarized in Fig.6, where the SRT and EGT predictions are compares with the experimental data.

\section{Conclusions and remarks}

An experimental and theoretical survey has been made on the one-way light propagation on Earth. The measurements have been obtained by using the TUPI diffraction laser experiment, and they have been analyzed on the basis of the EGT formalism. We find that there is evidences for an anisotropic light propagation (a first-order ether drift effect), because an amplified Doppler shift has been observed through the overlap of diffraction images as a function of the laser beam alignment relative to the Earth's velocity vector. This behavior arises of the Earth's motion (solar barycenter) relative to the CMBR rest frame. 

These experimental results contradict the so called Lorentz's theorem \cite{lorentz}, which states that the course of the relative rays is not affected by the motion of the Earth if quantities of the second order are neglected.
Now, after our experimental results, that certainly will be confirmed by other similar experiments, 
we conclude that the course of the rays is affected by the motion of the Earth, and a predominant quantity of first order describes well the experimental results. Of course, the observation of this effect requires a careful determination of the ``ether drift direction'', and only in 1991 after the publication of the COBE's results, that we know this direction with accuracy.

Our results point out that is not possible to neglect the preferred frame imposed by the cosmology. These results are well described by the Ether Gauge Theory (EGT) and which incorporates the cosmological time constraint.  
However, when only local aspects are considered the EGT predictions agree with the SRT predictions, they are basically
indistinguishable locally. 

Finally, we are conscious that only after an independent confirmation of our experimental data that they will have a total credibility and we hope that this happens soon, because the experiment is easy to reproduce.
Until then, we have the conviction that the universe behaves as a birefringent crystal ball.

\section{Acknowledments}

This paper is a memorial tribute to our professor and friend C.M.G. Lattes, he was who introduced us the non-conventional aspects about relativity. We are very grateful to K.H. Tsui for his assistance in preparation of the manuscript.

%%%%%%%%%%%%%%%%%%%%%%%%%%%%%%%%%%%%%%%%%%%%%%%%%%%%%%%%%%%%%%%%%%%%%%%%%%%%%%%%%%%%%%%% 
\newpage 

\begin{figure}[th]
\vspace*{-7.0cm}
\includegraphics[clip,width=1.0
\textwidth,height=1.0\textheight,angle=0.] {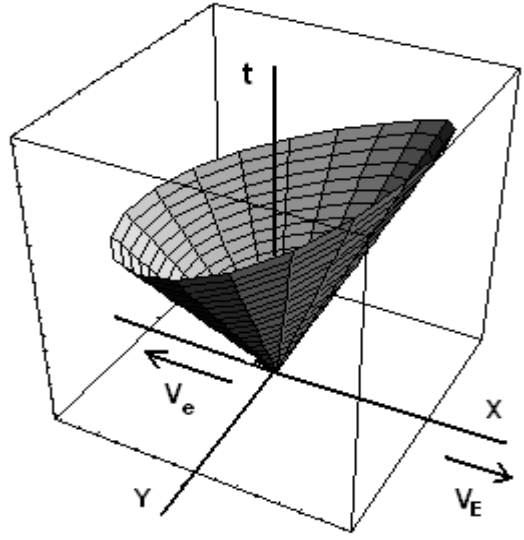}
\vspace*{-9.0cm}
\caption{\textbf{Space-time diagram (light cone) in the $\Sigma_E$ frame according to the EGT formalism}. It is possible to see that the axis-X  (direction of $V_E$) behaves as a preferred direction.  The slope of the lines are proportional to the inverse value of the speed of light. $V_e$ is the resulting vector speed of light (emission vector).  For illustrative purpose, a large value for $\beta(=0.3)$ in this diagram has been used. The effect for instance, due to the Earth motion relative to the CMBR rest frame is tiny ($\beta=0.001237$).}%
\end{figure}

\begin{figure}[th]
\vspace*{-2.0cm}
\hspace*{-2.0cm}
\includegraphics[clip,width=0.7
\textwidth,height=0.7\textheight,angle=0.] {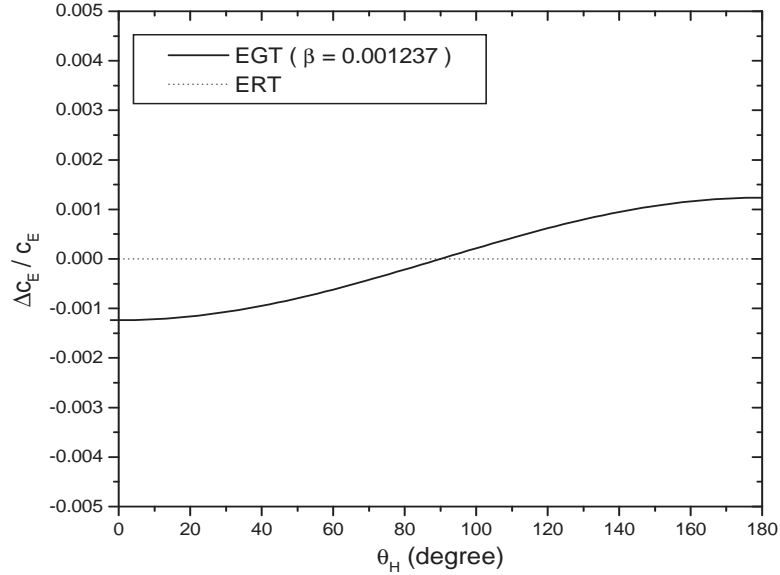}
\vspace*{-7.0cm}
\caption{\textbf{Angular dependence relative to the Earth's velocity vector of the ``dipole'' speed of light}. The lines represent the expected values according to the EGT ($\beta=0.001237$) and SRT formalisms.}%
\end{figure}

\begin{figure}[th]
\vspace*{-10.0cm}
\hspace*{-3.0cm}
\includegraphics[clip,width=1.2
\textwidth,height=1.2\textheight,angle=0.] {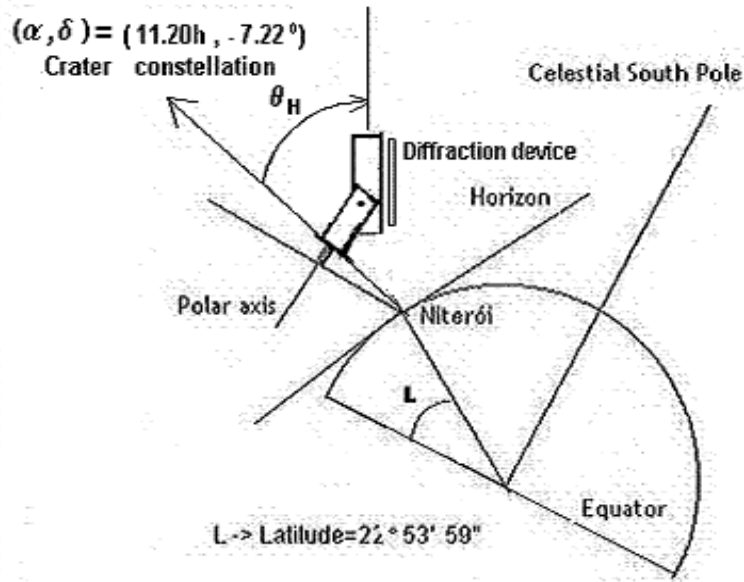}
\vspace*{-10.0cm}
\caption{\textbf{The equatorial ensemble of the TUPI muon telescope (raster scan system)}. The axis of the telescope can be pointing for a pre-established direction. The diffraction device is mounted in the shell of the telescope. The laser beam is parallel to the telescope axis. For $\theta_H=0$ the telescope points to the Crater constellation.
}%
\end{figure}

\begin{figure}[th]
\vspace*{-6.0cm}
\hspace*{-2.0cm}
\includegraphics[clip,width=0.9
\textwidth,height=0.9\textheight,angle=0.] {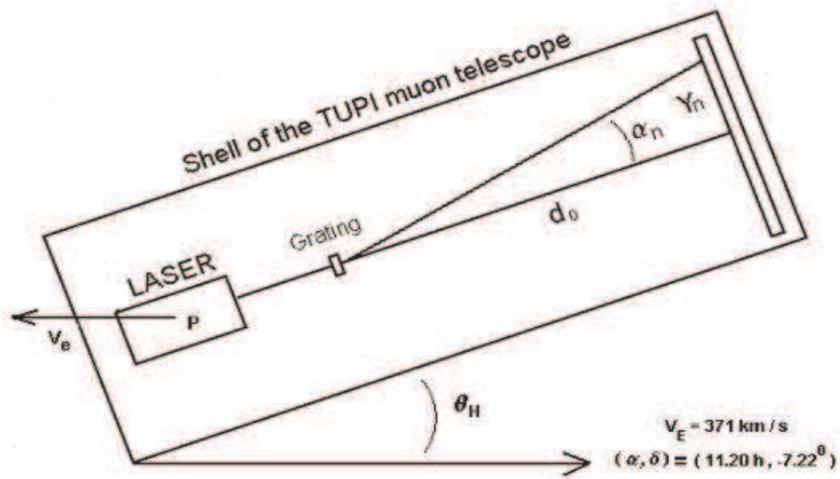}
\vspace*{-7.0cm}
\caption{\textbf{Schematic representation of the TUPI laser diffraction experiment in the Earth ($\Sigma_E$ frame)}. The emission vector is represented as $V_e$. }%
\end{figure}

\begin{figure}[th]
\vspace*{-5.0cm}
\hspace*{-1.0cm}
\includegraphics[clip,width=0.9
\textwidth,height=0.9\textheight,angle=0.] {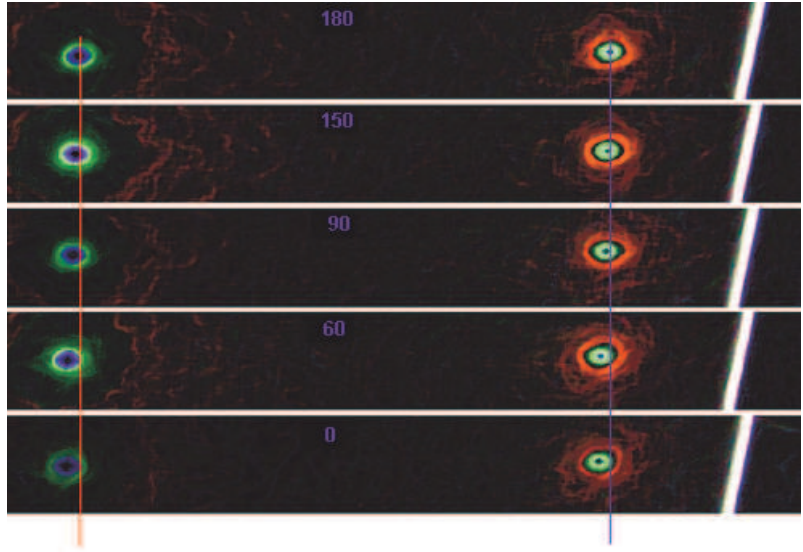}
\vspace*{-7.0cm}
\caption{\textbf{  
Amplified Doppler shift observed through the overlap of diffraction images as a function of the laser beam alignment relative to the Earth's velocity vector}. The images were filtered and amplified using always the same criteria. The colors in the images are artificial and the vertical lines were placed only as a visual guide.}%
\end{figure}

\begin{figure}[th]
\vspace*{-4.0cm}
\hspace*{-1.0cm}
\includegraphics[clip,width=0.7
\textwidth,height=0.7\textheight,angle=0.] {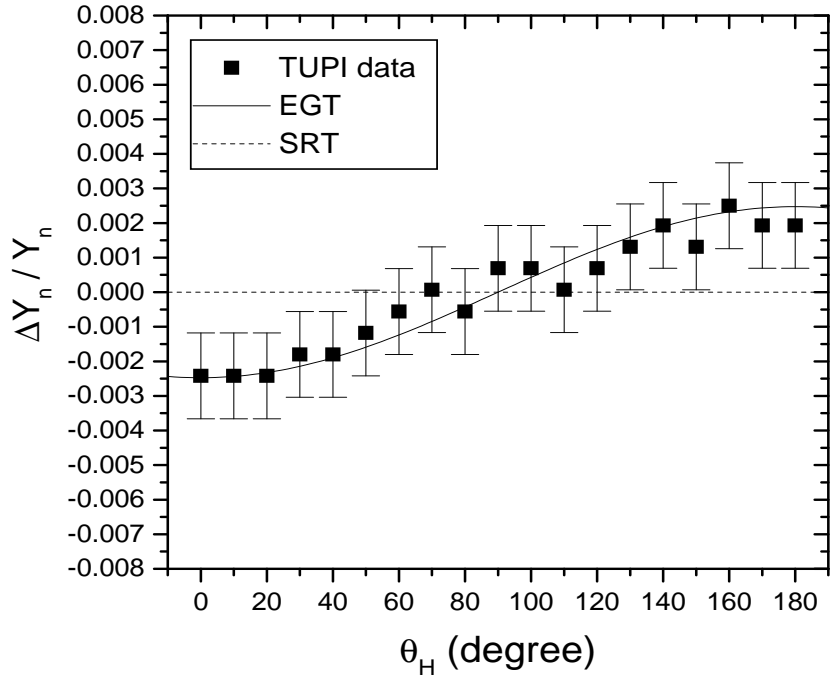}
\vspace*{-5.0cm}
\caption{\textbf{Angular dependence relative to the earth vector velocity of the relative amplified Doppler shift of  the diffraction images}. The expected curves according to the SRT and EGT($\beta=0.001237$) and experimental data are shown.}%
\end{figure}

\end{document}